\documentclass[numerical,10pt,superscriptaddress]{revtex4-1}
\usepackage{graphicx}
\usepackage{amsmath,amssymb}
\usepackage{times,mathptmx}
\usepackage{units}
\usepackage{epsfig}
\usepackage{booktabs}
\usepackage{multirow}

\usepackage[latin1]{inputenc}
\usepackage[english]{babel}
\usepackage{color}
\usepackage[symbol]{footmisc}

\newcommand{\VUB}{Applied Physics Research Group, Vrije Universiteit Brussel, Pleinlaan 2, 1050 Brussels, Belgium}

\newcommand{\HP}{Hewlett Packard Labs, 940 North McCarthy Blvd, Milpitas, CA 95035, USA}

\begin{document}

\date{\today}
\title{Order-of-magnitude differences in computational performance of analog Ising machines induced by the choice of nonlinearity}

\author{Fabian B{\"o}hm}\email[Corresponding authors: ]{Fabian B\"ohm (fabian.bohm@vub.be), Guy Van der Sande (Guy.Van.der.Sande@vub.be)}\affiliation{\VUB}
\author{Thomas Van Vaerenbergh}\affiliation{\HP}
\author{Guy Verschaffelt}\affiliation{\VUB}
\author{Guy Van der Sande}\email[Corresponding authors: ]{Fabian B\"ohm (fabian.bohm@vub.be), Guy Van der Sande (Guy.Van.der.Sande@vub.be)} \affiliation{\VUB}


\begin{abstract}

\section*{Abstract}
Ising machines based on nonlinear analog systems are a promising method to accelerate computation of NP-hard optimization problems. Yet, their analog nature is also causing amplitude inhomogeneity which can deteriorate the ability to find optimal solutions. Here, we investigate how the system's nonlinear transfer function can mitigate amplitude inhomogeneity and improve computational performance. By simulating Ising machines with polynomial, periodic, sigmoid and clipped transfer functions and benchmarking them with MaxCut optimization problems, we find the choice of transfer function to have a significant influence on the calculation time and solution quality. For periodic, sigmoid and clipped transfer functions, we report order-of-magnitude improvements in the time-to-solution compared to conventional polynomial models, which we link to the suppression of amplitude inhomogeneity induced by saturation of the transfer function. This provides insights into the suitability of systems for building Ising machines and presents an efficient way for overcoming performance limitations.

\end{abstract}

\maketitle

\section*{Introduction}

With the on-setting end of Moore's law, we also start to see an end to the continuous growth of both performance and energy efficiency of conventional von-Neumann-based digital computers \cite{THE17}, which is creating a particular challenge for performance and energy intensive tasks such as optimization and machine learning \cite{STR19,AND15}. Based on the well-known Ising spin model, Ising machines have emerged as a promising non-von-Neumann computing scheme that can accelerate computation of NP-hard optimization problems compared to conventional digital computers \cite{JOH11,YAM17a}. By mapping the cost function of optimization problems to an Ising Hamiltonian and implementing this Hamiltonian with a physical spin systems, calculation of optimal solutions can be achieved by the natural tendency of the spin system to evolve to its lowest energy state. Compared to conventional optimization algorithms such as simulated annealing, this natural analog computing concept can yield faster calculation and better energy efficiency \cite{DEN16,INA16a, HAR16b}. Various types of Ising machines have been proposed based on optical, electronic and quantum systems \cite{JOH11,SAL15,BAR16,KIM10,YAM16,CHO19}. Optical systems in particular have attracted great attention due to their high analog bandwidth, favorable energy dissipation and inherent parallelism \cite{UTS11,SHO17,BAB19,WAN13,INA16a,MCM16,BOE19,BER17,TEZ20,PIE19,PRA20,OKA20}. Such systems have also been adapted into different physics-inspired algorithms that have demonstrated equal performance with state-of-the-art optimization algorithms \cite{LEL19,KAL18,TIU19,GOT19}. A common feature among many of these Ising machines is that they are gain-dissipative nonlinear systems. Gain-dissipative systems generate spin states through a bifurcation-induced bistability that results from the interplay of linear gain dynamics with a nonlinear transfer function \cite{LEL17}. Originally, nonlinear systems based on supercritical pitchfork bifurcations have been proposed, as they naturally incorporate the Ising model and have demonstrated efficient ground states calculations for a variety of different optimization problems \cite{LEL17,LEL19,KAL18,BOE18,INA16}. However, numerous Ising machine designs have since then demonstrated that a larger variety of differing nonlinear systems can be used to implement Ising machines \cite{CHO19,BOE19,TIU19}. 

This raises the fundamental question what type of general nonlinear systems are capable of implementing Ising machines and which one is most suitable to achieve high computational performance in finding optimal solutions, i.e. short time-to-solution and high solution quality. A direct comparison between different Ising machines can be quite challenging though, due to the large differences in analog bandwidth and stability between different designs. Such engineering challenges have lead to differing claims about the advantages of particular systems \cite{BOE19,TIU19,CHO19}, while little insight has been gained thus far into what features make a general nonlinear dynamical system suitable as an Ising machine. This is of particular interest since the analog nature of the spin system typically results in amplitude inhomogeneity, which is known to lead to an incorrect mapping of the spin system to the target Ising Hamiltonian and thus inhibits the ability to find optimal solutions \cite{LEL17}. While active feedback systems have been proposed to counteract this inhomogeneity \cite{KAL18, LEL19}, such systems require to dynamically control the gain of each individual spin, which creates a significant overhead and could negatively affect the analog bandwidth. 

In order to effectively enhance the computational performance of analog spin systems, we consider the choice of the Ising machine's nonlinear transfer function as an efficient way of mitigating amplitude inhomogeneity. By unifying different Ising machine concepts into a generalized nonlinear dynamical system that makes their computational performance directly comparable, we identify general features in the system's nonlinear transfer function required for the implementation of Ising spins. Based on this, we simulate Ising machines with polynomial, periodic, sigmoid and clipped functions. To understand the influence of the nonlinear transfer function on the computational performance of Ising machines, we perform various benchmarks of the different nonlinearities based on NP-hard MaxCut optimization problems. We find that, while conventional systems based on pitchfork normal forms are often unable to find optimal solutions due to amplitude inhomogeneity, clipped and sigmoid nonlinear transfer functions can reach higher solution qualities and yield order-of-magnitudes improvements in the time-to-solution for the same problems. We link this enhanced computational performance to the strong suppression of amplitude inhomogeneity by the nonlinear transfer function, which shows that errors induced by the analog system can in part be compensated by choosing an appropriate nonlinear system. Our findings propose a straightforward and efficient way for overcoming computational performance deterioration due to amplitude inhomogeneity and motivate that a much larger variety of physical system beyond the current state-of-the-art can be considered for future generations of Ising machines. 

\section*{Results}

\subsection*{Generalized Ising machine model}

\begin{figure}[htbp]
	\centering
		\includegraphics[width=.49\textwidth]{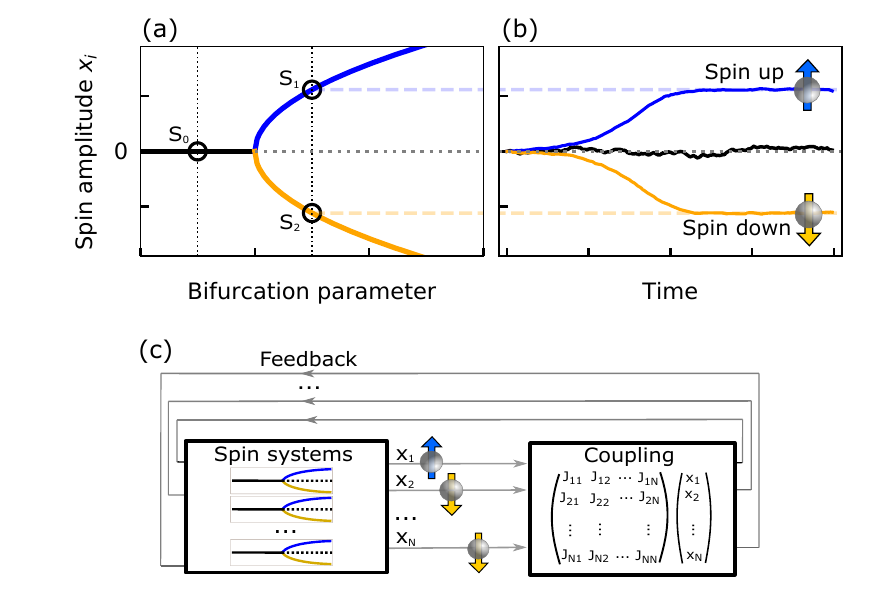}
	\caption{\textbf{Schematic of analog Ising spin systems} (a) Bifurcation diagram of a single gain-dissipative system as a function of the bifurcation parameter. Unstable fixed points are indicated by a dotted line. (b) Time evolution of a gain-dissipative system below the bifurcation point (black line) and above the bifurcation into the spin up (blue line) and spin down state (orange line) respectively. (c) Conceptual design of coupled gain-dissipative feedback systems to form an Ising machine. Spin states are generated in parallel gain-dissipative systems and coupled according to the coupling topology $J_{ij}$. The states are then fed back to the gain-dissipative systems to close the feedback loop.}
	\label{fig1}
\end{figure}

Ising machines are physical systems that implement coupled binary spins $\sigma_i=\left\{ -1,1 \right\}$, so that their energy or gain are equivalent to the Ising Hamiltonian

\begin{equation}
	H_{\mathrm{Ising}}=-\frac{1}{2}\sum_{ij}^N J_{ij}\sigma_i \sigma_j - \sum_i^N b_i \sigma_i \ \ . \label{eq1}
\end{equation}

The spins are either in the spin up ($\sigma_i=1$) or spin down state ($\sigma_i=-1$) and are coupled through the symmetric spin coupling matrix $J_{ij}$. Additionally, biases $b_i$ can be applied to any spin. The computational capabilities of the Ising machines arise from the fact that the cost function of various NP-hard combinatorial optimization problems can be directly mapped to such an Ising Hamiltonian \cite{LUC14} in a way that optimal solutions correspond to global energy minima of eq.\eqref{eq1}. The natural tendency of Ising machines to evolve to their lowest energy configuration is then used to find optimal solutions. A crucial challenge in building Ising machines is to find physical systems with a high analog bandwidth that can implement large networks of spins. A common way to achieve this is by using gain-dissipative systems. These are nonlinear systems with an analog spin variable $x_i$ that exhibit a bifurcation structure with a symmetrical bistability. 

Figure \ref{fig1}a shows a typical bifurcation diagram of a bistable gain-dissipative system as a function of the bifurcation parameter. Below the bifurcation point, the system has only one stable fixed point $S_0$ with an amplitude of $x_i(S_0)=0$. Above the bifurcation point, this trivial fixed point becomes unstable and two new fixed points $S_1$ and $S_2$ emerge that lie symmetrically around $S_0$. Figure \ref{fig1}b shows the time evolution of the spin amplitude $x_i$ when it is initially in the fixed point $S_0$. When the system is below the bifurcation point (black curve), the amplitude $x_i$ is fluctuating around the trivial fixed point $S_0$ due to the inherent noise of the system. Above the bifurcation point (orange and blue trace), the trivial fixed point becomes an unstable saddle, so that $x_i$ will either grow or decrease away from $S_0$ until it ends up in one of the fixed points $S_1$ or $S_2$. This binary nature is exploited to implement the Ising model. By extracting the sign of the spin amplitude, $x_i$ can be mapped to an Ising spin through $\sigma_i=sign(x_i)$. 

To implement the Ising Hamiltonian, Ising machines couple several of such analog Ising spins together. Figure \ref{fig1}c shows a schematic view of an Ising machine. Typically, an Ising machine is a continuous feedback system, where several bistable gain-dissipative systems are coupled with each other according to the spin coupling matrix $J_{ij}$. The dynamics of such a feedback system can then be modeled by the dimensionless differential equation
\begin{equation}
	\frac{dx_i}{dt}=F_i\left[x_i(t-\tau), \alpha, \beta\sum_j{J_{ij}x_j(t-\tau), \gamma\zeta_i(t)} \right] \ \ . \label{eq2}
\end{equation}
Here, $F$ is the nonlinear transfer function of the gain-dissipative systems and $\alpha$ and $\tau$ are the linear gain and the time delay of the feedback loop. The coupling between different spins occurs with the coupling strength $\beta$. To model noise, a Gaussian white noise term $\gamma\zeta$ is introduced with a zero mean and a standard deviation of $\gamma$. For simplicity, we neglect the time delay $\tau$ in the following. For optical and analog electronic systems in particular, this is often a reasonable assumption due to the short time of flight of light. The central questions that we are addressing in this work is how the nonlinear transfer function $F$ has to be chosen in order to be suitable for Ising machines and how the particular choice of a nonlinear system affects the computational performance when solving optimization problems. In the following, we show how suitable dynamical systems can be constructed from general classes of nonlinear functions, namely polynomial, periodic, sigmoid and clipped functions and we compare the computational performance of these different nonlinearities.

\subsection*{Ising machines based on polynomial functions}

\begin{figure}[htbp]
	\centering
		\includegraphics[width=.99\textwidth]{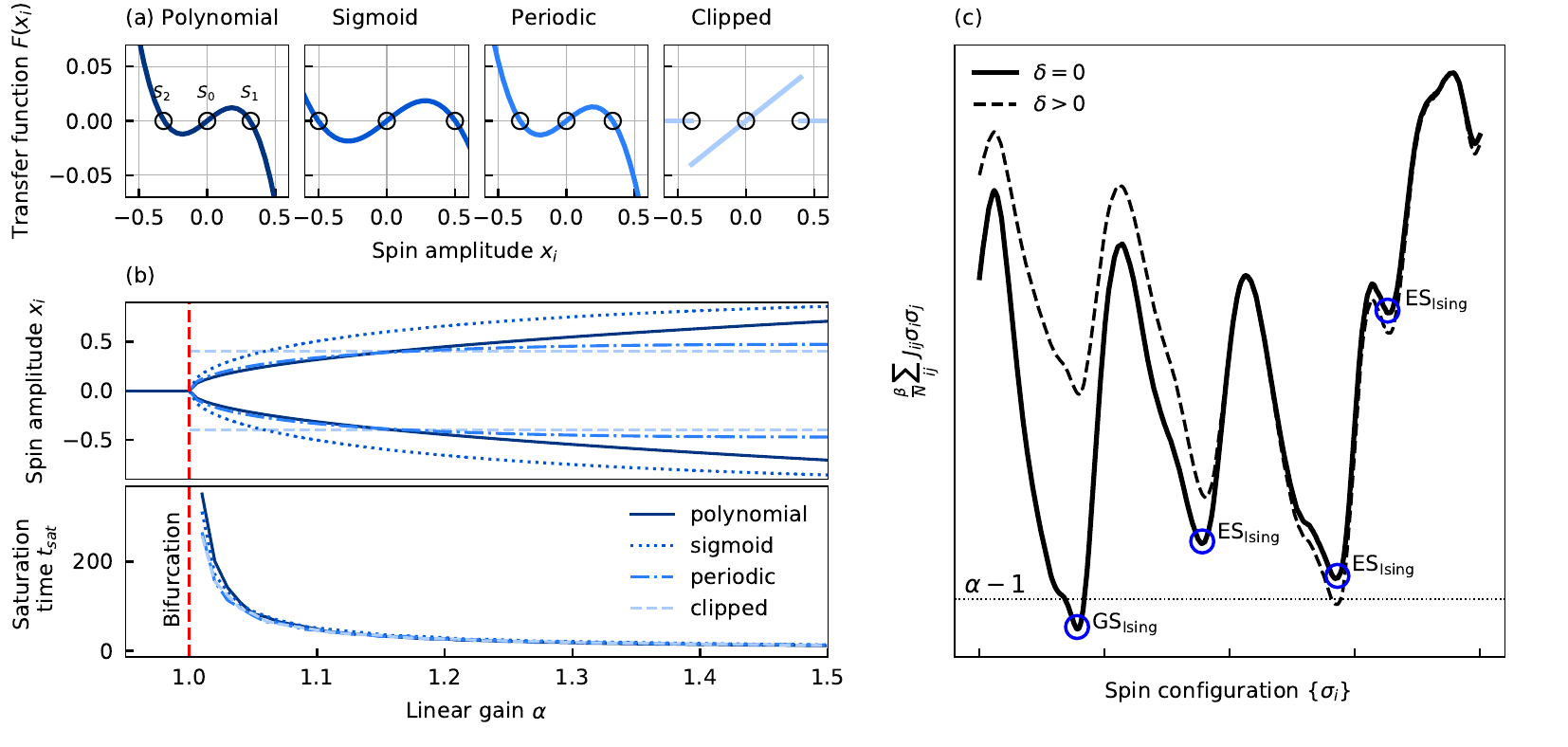}
	\caption{\textbf{Implementation of analog Ising spins with different nonlinear transfer functions} (a) Nonlinear transfer function as a function of the spin amplitude for polynomial, sigmoid, periodic and clipped nonlinearities for isolated spins. (b) Bifurcation diagram and average saturation time for isolated spins for the nonlinearities in (a) as a function of the linear gain $\alpha$. (c) Visualization of the inequality\eqref{eqgain} for the case of homogenous spin amplitudes ($\delta=0$, solid line) and inhomogeneous amplitudes ($\delta > 0$, dashed line). The dotted line indicates the left-hand-side of eq.\eqref{eqgain} for an arbitrary $\alpha$. Fixed points corresponding to global minima (ground states $\mathrm{GS_{Ising}}$) and local energy minima (excited states $\mathrm{ES_{Ising}}$) of the Ising model are indicated by circles. For homogeneous amplitudes, fixed points that lie below the dotted line fulfill the condition of eq.\eqref{eqgain} and therefore exist.}
	\label{fig2}
\end{figure}

A basic way to generate an Ising spin system with polynomial transfer functions is the pitchfork normal form. The pitchfork normal form is inherent in various optical systems and has been used to describe Ising machines, e.g. for the classical approximation of degenerate optical parametric oscillators \cite{WAN13,HAR16b,OKA20}, Kerr-nonlinear microring resonators \cite{TEZ20} and polariton condensates \cite{KAL18}. The nonlinear transfer function of Ising machines based on the supercritical pitchfork normal form is given by (arguments of $F$ have been omitted for clarity)

\begin{equation}
	F_i(\{x_i\})= (\alpha-1) x_i -x_i^3 + \beta\sum_j{J_{ij}x_j} + \gamma\zeta_i(t)  \label{eq4}
\end{equation}

and consists of a linear growth term with the linear gain $\alpha$, a cubic saturation term and a coupling term with coupling strength $\beta$. In the following, we first consider the dynamics of the uncoupled system ($\beta=0$). Figure \ref{fig2}a shows the right hand side of eq.\eqref{eq4} at $\alpha=1.1$ for an isolated spin ($\beta=0$) as a function of the spin amplitude $x_i$. Characteristically, the transfer function contains three zero crossings, which correspond to three fixed points. $S_0$ is at the origin and $S_1$ and $S_2$ symmetrically surround the origin. In between the fixed points, there are a local minimum to the left and a local maximum to the right of the origin, which results in an S-shaped transfer function. From linear stability analysis, it follows that the central fixed point is unstable, while the two surrounding fixed points are bistable. 

The corresponding bifurcation diagram resulting from this transfer function is shown in the top panel of fig.\ref{fig2}b. The uncoupled system possesses a pitchfork bifurcation with the bifurcation point at $\alpha=1$ (indicated by the red dashed line). Below the bifurcation point, only the trivial fixed point $S_0$ is stable. Above the bifurcation point, the trivial solution becomes unstable and the two symmetrically bistable fixed points $S_1$ and $S_2$ arise. The amplitude of $S_1$ and $S_2$ is growing monotonically with $\alpha$ and scales as $\left|S_{1,2}\right|\propto\sqrt{\alpha-1}$. In the bottom panel of fig.\ref{fig2}b, we consider the dynamical timescale of this system as a function of $\alpha$ by measuring the saturation time $t_{sat}$, i.e. the average time it takes the spin amplitude to grow/decrease to half of the fixed points' amplitude. Directly at the bifurcation point, we observe critical slowing down of the temporal evolution of $x_i(t)$, where the saturation time increases exponentially towards the bifurcation \cite{KUE11}. This feature of the bifurcation can easily be understood by considering the shape of the transfer function $F(x_i(t))$. As $\alpha$ approaches $\alpha=1$, the magnitude of the linear growth term in eq.\eqref{eq4} becomes vanishingly small so that the growth rate of the spin amplitude stagnates. 

To understand the computational capabilities of Ising machines, we now consider a network of pitchfork normal forms that are coupled according to the coupling matrix $J_{ij}$. The ability of eq.\eqref{eq4} to implement the Ising model \eqref{eq1} can be understood by deriving the Lyapunov function $L(\{x_i\})$. The Lyapunov function is a measure of the stability of a particular amplitude configuration $\{x_i\}$, where stable configurations correspond to minima of $L(\{x_i\})$. For the Ising machine, the Lyapunov function is obtained by integrating the equation of motion \eqref{eq4} and summing over all spins:
\begin{equation}
	L(\{x_i\})=-\sum_i\left((\alpha-1)\frac{x_i^2}{2}-\frac{x_i^4}{4} \right)-\frac{\beta}{2}\sum_{ij}J_{ij} x_i x_j \ \ . \label{eqlya}
\end{equation}
In the case of homogeneous spin amplitudes ($|x_i|=const.$), we find a direct correspondence of the Lyapunov function to the Ising model. While the first two terms are constant regardless of the amplitude configuration, the last term is formally equivalent to the Ising Hamiltonian \eqref{eq1}. By taking the sign of the spin amplitude $\sigma_i = \frac{|x_i|}{x_i}$, the Lyapunov function thus contains the same minima as the Ising model so that the ground state corresponds to the global minimum of $L(\{x_i\})$. Since by definition $\frac{dL}{dx_i}=-F(x_i)$, the minima are stable fixed points of the coupled system. As we detail in the methods section, the condition for every fixed point to exist for homogeneous amplitudes is given by \cite{LEL17} 
\begin{equation}
	\alpha-1 \geq \frac{\beta}{N}\sum_{ij}J_{ij} \sigma_i \sigma_j\ \ . \label{eqgain}
\end{equation}
This inequality is visualized in Fig.\ref{fig2}c. The solid line depicts the r.h.s of eq.\eqref{eqgain} as a function of the spin amplitude configuration $\{\sigma_i\}$ for an exemplary Lyapunov function. The l.h.s. of eq.\eqref{eqgain} for an arbitrary line gain $\alpha$ is indicated by the black dotted line. The inequality dictates that only fixed points corresponding to local minima below the dashed line exist. The Ising machine is therefore unable to reach any of the energy minima above the dotted line. This condition is exploited to effectively single out the ground state. As $\alpha$ increases from the region where only the trivial solution $\{x_i\}=0$ exists, the first non-trivial solution to exist is the ground state (GS), while suboptimal solutions are still nonexistent and can therefore be avoided. 

However, it is important to note that the assumption of homogeneous amplitudes cannot be made in general \cite{LEL17}. Due to their analog nature, the spin amplitudes are inhomogeneously distributed around the fixed points with a standard deviation $\delta$. As we show in the methods section, this leads to a modification of the coupling matrix $J_{ij}$ of the implemented Ising model so that it no longer corresponds to the intended target Hamiltonian \eqref{eq1}. The dashed line in Fig.\ref{fig2}c exemplifies the influence of amplitude inhomogeneity ($\delta>0$) on the Lyapunov function. Compared to the homogeneous case ($\delta=0$), the inhomogeneity can induce a relative shift of the energy minima so that the ground state minimum is no longer the lowest energy configuration or is erased altogether. In this case, the lowest configuration corresponds to an excited state (ES), while the ground state can only be reached at much higher gain levels. This incorrect mapping of the target Ising model can thus significantly deteriorate or even diminish the probability of finding optimal solutions in optimization tasks. To mitigate this issue, a proposed approach is to modulate the gain of each individual spin individually to force the spins to one common amplitude \cite{KAL18,LEL19,KAN21}. However, this effectively doubles the number of dynamical equations and requires the addition of an active feedback system to perform the calculations of all gain coefficients, which can create a significant overhead and potentially reduce the analog bandwidth. In the following, we thus want to consider other types of nonlinearities and investigate how they can be used to directly reduce the negative effect of amplitude inhomogeneity.

\subsection*{Ising machines based on sigmoid functions}

From the polynomial model \eqref{eq4}, we find that the shape of the nonlinear transfer function is essential for generating analog Ising spins. Here, we investigate how such spin systems can be generated by mimicking the shape of the transfer function \eqref{eq4} with sigmoid functions. While sigmoid functions have so far not been considered for Ising machines, they are widely used in the context of Hopfield-Tank-networks and other neuromorphic systems to mimic the activation function of neurons \cite{HOP85}. Efficient ways of implementing them have been reported for both optical systems and electronic systems \cite{MIS18,WIL20,JHA20,MOU19,LU00,TSM19}. Sigmoid functions are characterized by an S-shaped nonlinearity and can be modeled by a variety of functions such as the logistic function or the Gompertz function. Here, we consider a sigmoid transfer function based on the hyperbolic tangent function

\begin{equation}
	F_i(\{x_i\})=-x_i+\tanh(\alpha x_i + \beta\sum_j{J_{ij}x_j} + \gamma\zeta_i(t)) \ \ . \label{eq5}
\end{equation}

To facilitate a simple comparison to the polynomial model, we expand eq.\eqref{eq5} into a Taylor series to the third order for small spin amplitudes. As we derive in the methods section, in the weak coupling regime $\alpha \gg \beta$, this results in $F(x_i)\approx (\alpha-1)x_i - \frac{\alpha^3 x_i^3}{3} + \beta\sum_j{J_{ij}x_j}$. Compared to the polynomial model, we recognize a close resemblance to eq.\eqref{eq4} with the same linear and nonlinear terms in $x_i$ as well as a linear coupling term. This suggests that the sigmoid model works as an approximation of the polynomial model when the system is close to the bifurcation point. We first consider the ability of the sigmoid model to implement uncoupled Ising spins ($\beta=0$). When comparing the shape of this transfer function for an isolated spin in Fig.\ref{fig2}a for $\alpha=1.1$ to that of the polynomial model, we find close similarities in its shape and in the position of the fixed points. In the bifurcation diagram in Fig.\ref{fig2}b, we observe the same bifurcation point at $\alpha=1$ and a good agreement of the fixed points for $\alpha \approx 1$. For higher linear gain, the amplitude of the fixed points starts to deviate due to the different coefficient in the third order polynomial term and due to the additional higher order terms. Particularly, the absolute amplitude of $|S_{1,2}|$ does not increase continuously but rather saturates for large $\alpha$ at $|S_{1,2}|\rightarrow 1$. Despite this, the saturation time in Fig.\ref{fig2}b agrees well with the polynomial model for all $\alpha$. This is to be expected, since the saturation time $t_{sat}$ primarily depends on the linear growth term, which is identical between both models. When the spins are coupled, the linear coupling term ensures a good approximation of the Lyapunov function in eq.\eqref{eqlya} to the Ising model for small amplitudes. While the additional higher order terms can cause small deviations from the polynomial model, the linear coupling term remains dominant so that good mapping to the Ising model can be expected. Although the concept of using sigmoid functions has been considered in neural systems before \cite{HOP85}, its ability to generate a bistable bifurcation structure indicates that they are also inherently suitability for generating Ising machines.

\subsection*{Ising machines based on periodic functions}

Periodic transfer functions form another set of nonlinearities that can be efficiently implemented with optical and electrical systems \cite{BOE19,CHO19}. To generate an Ising spin system with periodic transfer functions, the general shape of the polynomial model \eqref{eq4} can be mimicked by appropriately shifting cosine or sine functions. In the following, we consider a nonlinear dynamical system based on a $cos^2$ nonlinearity

\begin{equation}
	F_i(\{x_i\})=-x +\cos^2\left(\alpha x_i-\frac{\pi}{4} + \beta\sum_j{J_{ij}x_j} + \gamma\zeta_i(t) \right)-\frac{1}{2} \ \ . \label{eq6}
\end{equation}

The $cos^2$ nonlinearity models Ising machines based on optical intensity modulators \cite{BOE19} but is also equivalent to electronic oscillator-based Ising machines \cite{CHO19}. As with the sigmoid model, we expand the transfer function in a Taylor series to the third order. For small amplitudes and for the weak coupling regime, this results in $F(x)\approx (\alpha-1)x-\frac{2\alpha^3x^3}{3} + \beta\sum_j{J_{ij}x_j}$. Similar to the sigmoid model, we find close resemblance to eq.\eqref{eq4}, which suggests a good approximation close to the bifurcation point. Comparing the transfer function for an isolated spin to that of the polynomial model in fig.\ref{fig2}a, we find that both systems closely resemble each other both in shape and in the position of the fixed points. In the bifurcation diagram in fig.\ref{fig2}b, we observe good agreement with the polynomial model for the amplitude of the fixed points when the system is close to the bifurcation point. For higher values of $\alpha$, the higher order terms and the different scaling with $\alpha$ causes deviations. As for the sigmoid model, the absolute amplitude of the fixed points does not increase continuously but rather saturates at around $|S_{1,2}| \rightarrow 0.5$. Still, the saturation time $t_{sat}$ is identical to that of the polynomial model over the entire range of $\alpha$ in fig.\ref{fig2}b, which is expected due to the matching linear growth term. For the coupled system, the linear coupling term ensures the correspondence of the Lyapunov function to the Ising model.

\subsection*{Ising machines based on clipped functions}

As a last class of functions, we consider transfer functions that are clipped. Clipping is inherent in various electronic systems due to load limitations of components and has for example been observed in opto-electronic Ising machines \cite{BOE19}. Clipping has also been proposed as an efficient way to emulate Ising machines with digital hardware \cite{TIU19}. In the following, we consider a linear transfer function that is clipped to a maximum value of $\left|x_i\right| \leq 0.4$:

\begin{equation}
    F_i(\{x_i\}) = 
		\begin{cases}
			(\alpha-1)x_i + \beta\sum_j{J_{ij}x_j} + \gamma\zeta(t),\ \text{for } \left|x_i\right|\leq 0.4\\
			0,\ \ \ \ \ \ \ \ \ \ \ \ \ \ \ \ \text{for } \left|x_i\right|> 0.4
		\end{cases}
		\label{eq7}
\end{equation}

While the transfer function contains the same linear growth term and coupling term as the polynomial model, clipping is quite different from the nonlinear saturation terms discussed before. For isolated spins, the transfer function depicted in fig.\ref{fig2}a only possesses one zero crossing and therefore only one fixed point. The function is discontinuous at the clipping levels with a sudden jump to zero beyond the clipping level, which pins the spins to the clipping level at $|S_{1,2}|=0.4$. This difference is clearly reflected in the bifurcation diagram in fig.\ref{fig2}b. Although the clipped function possesses the same bifurcation point at $\alpha=1$ as the previous models for isolated spins, the amplitude of the fixed points does not increase or decrease with $\alpha$, but rather immediately jumps to the clipping level at the bifurcation point. Hence, while the linear growth term and therefore the growth rate of the spin amplitude may be similar to the other models, the saturation amplitude can be quite different at the bifurcation. Setting the level to 0.4 therefore ensures that the spin amplitude remains comparable to the other models at a gain close to the bifurcation point. As a consequence, the saturation time $t_{sat}$ in fig.\ref{fig2}b remains very similar to that of the other models with the same critical slowing down at the bifurcation point. For the coupled system, the clipped model possesses the same linear coupling term as the polynomial model. Contrary to the sigmoid and periodic models, no additional nonlinear coupling terms are contained in the transfer function, which ensures the ability to implement the Ising model for homogeneous amplitudes.

\subsection*{Parameter optimization for inhomogeneous spin amplitudes}

\begin{figure}[htbp]
	\centering
		\includegraphics[width=.99\textwidth]{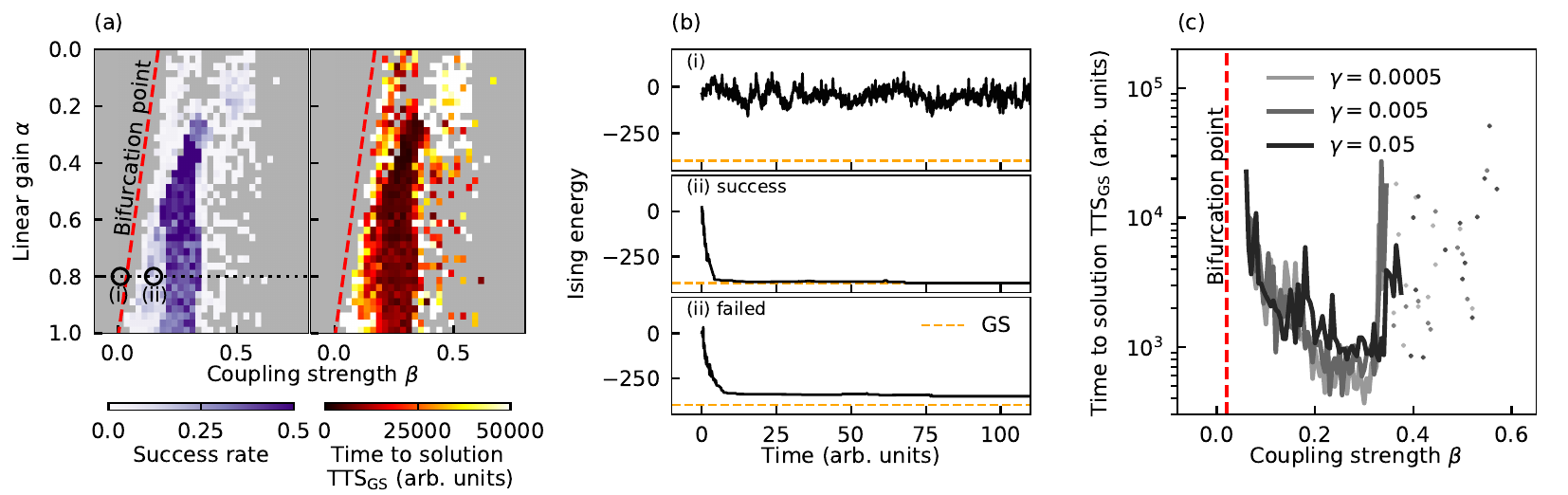}
	\caption{\textbf{Influence of parameters on computational performance} (a) Success rate and time-to-solution for the random graph $g05_{100,5}$ for a scan of $\alpha$ and $\beta$. The pitchfork bifurcation point is indicated by the dashed line. The grey region indicates the points where the Ising machine is unable to converge to the ground state. (b) Exemplary time evolution of the Ising energy below the bifurcation point (top panel, corresponds to (i) in a)) at $\alpha=0.01$ and above the bifurcation point at $\alpha=0.15$ for successful (middle panel, corresponds to (ii) in a)) and unsuccessful (bottom panel) convergence to the ground state (GS). (c) Time-to-solution as a function of the coupling strength $\beta$ for different noise $\gamma$ strengths at $\alpha=0.8$. In (a) and (b), $\gamma$ is fixed at $\gamma=0.005$. }
	\label{fig3}
\end{figure}

As described in the previous section, finding ground states with analog Ising machines follows the same approach for arbitrary Ising models in the case of homogenous spin amplitudes. Indeed, for homogeneous amplitudes, the ground state is the first and only solution to exist and can be reached by finding the bifurcation point, either by gradually increasing the gain or by analytical methods \cite{WAN13,LEL17}. However, this simple scheme fails in general due to amplitude inhomogeneity. With amplitude inhomogeneity, the ground state may not always exist directly at the bifurcation point, which requires to scan the gain above the bifurcation point until the condition for existence is fulfilled. Furthermore, the ground state can become multistable with other excited states, which makes the ground state search non-deterministic and requires to find operating regions with higher probability to find the ground state. 

In order to optimize the performance of the Ising machines using the different transfer functions, we perform scans of the linear gain $\alpha$, the coupling strength $\beta$ and the noise strength $\gamma$. We optimize the performance in regard to the success rate $P_a$ as well as the time-to-solution $\mathrm{TTS}_a$. The success rate $P_a$ measures the probability of reaching a specific solution $a$ (e.g. the ground state) at any point after initializing the Ising machine. The time-to-solution, defined as 
\begin{equation}
	\mathrm{TTS}_a=T_{a}\frac{\log(0.01)}{\log(1-P_{a})} \label{eq8} \ ,
\end{equation}
measures the time required to reach the solution $a$ with 99 percent probability. It is calculated from the success rate $P_a$ and the average time $T_a$ to reach that solution. $T_a$ is calculated by tracking the energy during the evolution of the Ising machine and corresponds to the point where the solution $a$ is first reached, either by converging or by a transient state. In figure \ref{fig3}a, we show exemplary success rates and time-to-solutions for reaching the ground state with the periodic model for a sweep of $\alpha$ and $\beta$. Here, $P_a$ is estimated by repeatedly initializing the Ising machine and counting the number of instances in which the ground state has been reached. The implemented Ising model is the random graph $g05_{100,5}$ contained in the Biq Mac graph library, which has a known ground state at $E_\mathrm{Ising}=-397$ \cite{REN07}. We have estimated the bifurcation point of this graph from the point where the trivial solution becomes unstable, which is indicated in the parameter space by the red dashed line. Compared to the case of isolated spins in fig.\ref{fig2}b, where the bifurcation point is at $\alpha=1$, the bifurcation point is shifted due to the coupling to the other spins. If $\beta$ is below the bifurcation point, the success rate is zero as the system is unable to bifurcate and no solution besides the trivial one exists. The corresponding time-to-solution is thus undefined. The top panel of fig.\ref{fig3}b shows an exemplary time series of the Ising energy in this parameter region for $\alpha=0.8$ and $\beta=0.01$. At this point, the Ising energy randomly fluctuates around zero due to noise. 

As $\beta$ is increased to be directly above the bifurcation point, we observe that the success rate remains at zero. This indicates that the first solutions are excited states and that the ground state position is likely shifted due to amplitude inhomogeneity. Only for higher $\beta$, the success rate gradually increases and the fixed point corresponding to the ground state starts to exist. At this point, $P_a$ is at around 10 percent, which indicates multistability with various other excited states. The middle panel of fig.\ref{fig3}b shows an exemplary time evolution of the Ising energy for a successful calculation for $\alpha=0.8$ and $\beta=0.1$. As the system is initialized, the Ising energy immediately decreases until the system eventually converges to a stable configuration at the ground state energy after $t\approx 70$. As a comparison in fig.\ref{fig3}b, we show a case in which the Ising machine reaches an excited state instead. After initially decreasing, the Ising energy converges to an energy of $E_\mathrm{Ising}=-349$, which is only at 88 percent of the ground state. For higher $\beta$, the likelihood of this undesired convergence to excited states reduces and the success rate increases to around 40 percent. The corresponding time-to-solution decreases with this rising success rate from $\mathrm{TTS_{GS}} \approx 10000$ to an optimum of $\mathrm{TTS_{GS}} \approx 1000$, as fewer repeated runs of the Ising machine are required until the ground state is found.  Eventually, for very high $\beta$, it becomes impossible again to reach the ground state and the success rate becomes zero. fig.\ref{fig3}a signifies the sensitivity of Ising machine performance to changes in $\alpha$ and $\beta$. For the $g05_{100,5}$ graph, there is a clear gap between the bifurcation point and the region where the ground state can be found. Furthermore, the operating point with the lowest time-to-solution is for values of $\beta$ that are further away from the point where the ground state first starts to exist. For each nonlinear transfer function, a sweep of $\alpha$ and $\beta$ is therefore necessary to determine the optimal operating point.

We also consider the influence of the noise strength $\gamma$ on the overall performance. Contrary to recent high noise level proposals for Ising machines with discrete spin systems \cite{ROQ20,PIE20}, we choose a noise level that is much smaller than the amplitude of the fixed points $S_{1,2} \gg \gamma$, which corresponds to experimental realizations of analog Ising machines. This ensures that the noise is not strong enough to switch the configuration of individual spins and therefore guarantees that the Ising machine always converges to a stable configuration. The noise will therefore not directly influence the linear stability and the overall success rate. To assess whether the noise has any influence on the performance, we perform sweeps of $\gamma$ over two orders of magnitude. In fig.\ref{fig3}c, we measure the time-to-solution for different $\gamma$ as a function of $\beta$ for the $g05_{100,5}$ graph at $\alpha=0.8$. Due to the non-deterministic nature of the ground state search, fluctuations of the time-to-solution within a factor of 2 around the average are observed for all noise levels. Interestingly, although $\gamma$ is changed over two orders of magnitude, we cannot identify a clear change in the time-to-solution. We have verified this result for the different nonlinearities with various Ising models and parameter configurations and observe the same trend. We conclude that as long as the noise level is sufficiently small, we can assume that $\gamma$ has a neglectable influence on the overall computational performance. In all following simulations, we have therefore fixed $\gamma$ to a constant value of $\gamma=0.005$.

\subsection*{Benchmark of different nonlinearities}

\begin{figure}[htbp]
	\centering
		\includegraphics[width=.49\textwidth]{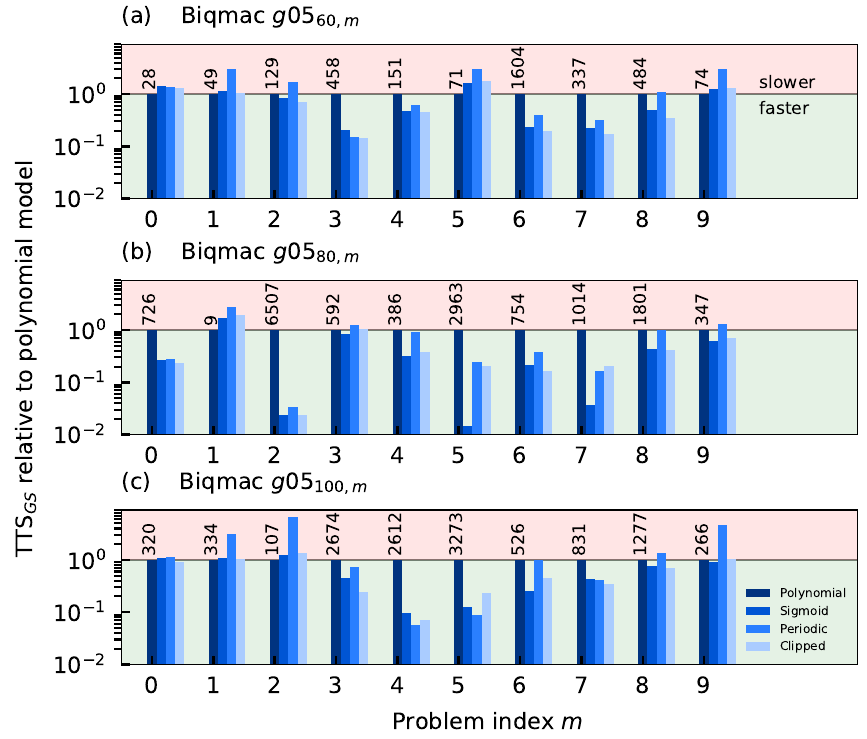}
	\caption{\textbf{Computational performance in Biq Mac benchmark tasks} Time-to-solution of the periodic, sigmoid and clipped model relative to the polynomial model for the Biq Mac g05 MaxCut benchmark set with $N=60$ (a), $N=80$ (b) and $N=100$ spins (c). The absolute value for the time-to-solution for the polynomial model is indicated by the numbers above the bars for the polynomial model.}
	\label{fig4}
\end{figure}

To consider the effect of the nonlinear transfer function on the computational performance of Ising machines, we benchmark the different systems with various MaxCut optimization problems. MaxCut is a task to maximize the cut number
\begin{equation}
	C=\frac{1}{4}\left(\sum_{ij}J_{ij}-\sum_{ij}^N J_{ij}\sigma_i \sigma_j \right) \ \  \label{eq9}
\end{equation}
when separating a graph structure into two parts and is known to be an NP-hard problem \cite{GAR90}. For the benchmarks, we use instances contained in the Biq Mac and the SuiteSparse Matrix Collection libraries. From the Biq Mac library, we consider the $g05_{N,m}$ subset of random undirected graphs with an edge density of 50 percent, for which the ground states are known \cite{REN07}. Similar to fig.\ref{fig3}a, we test all 10 different instances for $N=60$, $N=80$ and $N=100$ respectively by performing sweeps of $\alpha$ and $\beta$ and measuring the time-to-solution to reach the ground state $\mathrm{TTS_{GS}}$. In fig.\ref{fig4}, we consider the best time-to-solution that was achieved by the different nonlinearities during the scan of $\alpha$ and $\beta$. For the periodic, sigmoid and clipped system, the time-to-solution is shown as a ratio to the time-to-solution achieved by the polynomial model, whose absolute value is shown as a reference (absolute values for all models are given in the supplementary table 1). While all nonlinearities are able to converge to the ground state, we observe drastic differences for some specific problems. In these cases, the polynomial model typically performs worse than the other models with a time-to-solution that is one or two orders of magnitude slower, which is beyond the noise-induced fluctuations in fig.\ref{fig3}c. We find that spins in all models still evolve on very similar timescales (similar to the isolated spins in fig.\ref{fig2}b). However, we observe significantly lower success rates for the polynomial model that cause the large differences in performance.

\begin{figure}[htbp]
	\centering
		\includegraphics[width=.99\textwidth]{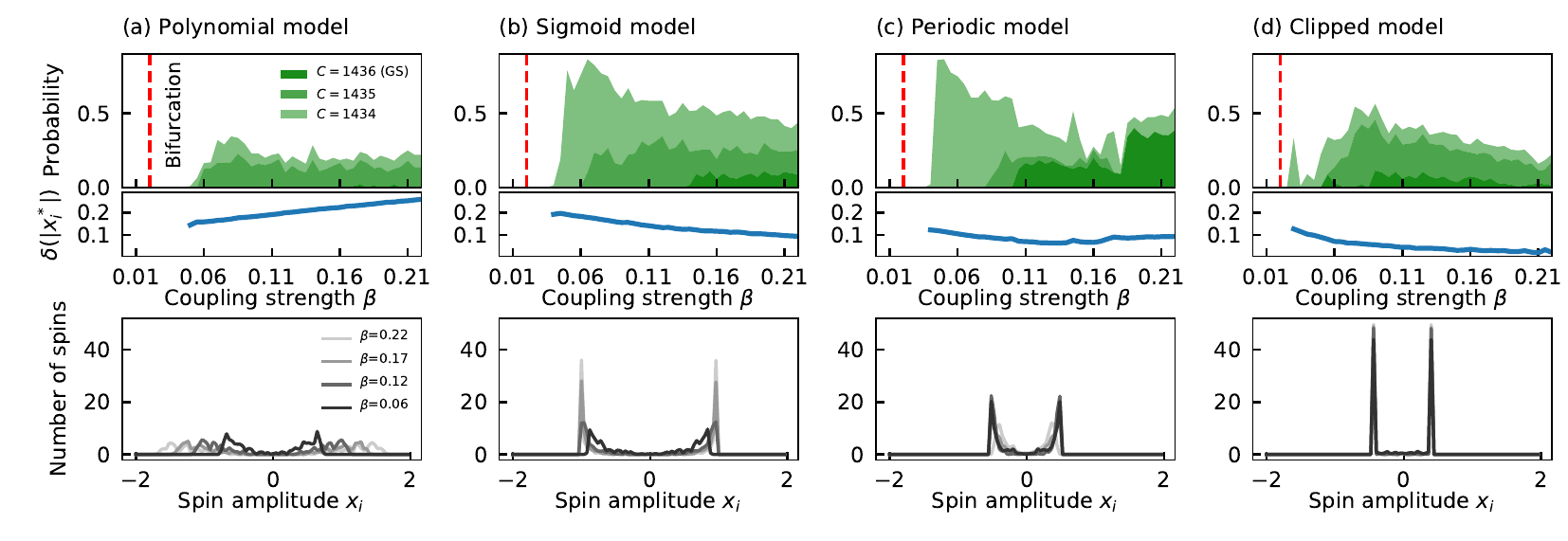}
	\caption{\textbf{Relation between amplitude inhomogeneity and success rate for different nonlinearities} Success rate (top), standard deviation $\delta$ of the fixed point (middle) and amplitude distribution (bottom) as a function of $\beta$ for the polynomial (a), sigmoid (b), periodic (c) and clipped model (d). The success rate shows the probability of reaching the three highest cut values with the ground state at $C=1436$. The standard deviation and the amplitude distribution have been calculated once the system reaches a steady state and are the average over the three highest cut values. The implemented MaxCut problem is $g05_{100,5}$ with $\alpha$ fixed at $\alpha=0.8$.}
	\label{fig7}
\end{figure}

To better understand these differences, we consider the Biq Mac instance $g05_{100,5}$ as an example, where the time-to-solution differs by around one order of magnitude between the polynomial model and the other nonlinearities. In fig.\ref{fig7}, we perform scans of the coupling strength $\beta$ through the parameter space from below to above the bifurcation and analyze the fixed points that the systems converge to. We select $\alpha=0.8$, as it corresponds to a region where all models have been able to find the ground state with success rates close to the optimum during the scans of $\alpha$ and $\beta$. In the top panels, we calculate the success rate to reach the three highest cut values for the different models (fig.\ref{fig7}(a)-(d)). We find that the polynomial model in fig.\ref{fig7}a is unable to reach the ground state at any point in the scan. We have verified this by initializing the polynomial model in the correct ground state configuration and observe that the system instead converges to excited states of the implemented target Ising Hamiltonian. We have also tested other instances in the $\alpha-\beta$ parameter scan, where the ground state was reached by the polynomial model. We have found that these instances are transient states that pass through the ground state before converging to an excited state. The probability of reaching the ground state through these transient states is at just 2 percent per run and thus significantly lower than the success rate of any of the other model, hence causing the high TTS in fig.\ref{fig4}. When considering the fixed points of the polynomial model, we are therefore unable to find any point in the parameter space at which the fixed point corresponding to the ground state exists. For the other models on the other hand, the ground state exists for increasing $\beta$ and the systems are all able to converge to the optimal solution at a much higher success rate. We therefore find that the nonlinear transfer function can considerably affect the ability to correctly implemented the desired Ising model.

Since the failure of the mapping to the target Ising Hamiltonian is typically associated with amplitude inhomogeneity of the fixed point \cite{LEL17}, we quantify the amount of inhomogeneity for the different models. We measure the standard deviation $\delta(|x^*_i|)$ of the absolute value of the amplitude $|x^*_i|$ for the fixed points corresponding to the three highest cut values. In the middle panel of fig.\ref{fig7}, we show $\delta(|x^*_i|)$ as a function of $\beta$ and compare it to the success rate in the upper row of fig.\ref{fig7} for each model (a-d). For the polynomial model, $\delta(|x^*_i|)$ continuously increases with $\beta$ and eventually doubles relative to the value at the bifurcation point. In the lower panel of fig.\ref{fig7}a, we show exemplary amplitude distributions for the polynomial model, which are smoothly and broadly distributed around the fixed points of the isolated spins in fig.\ref{fig2}b. We observe how the distribution becomes broader for high $\beta$ as amplitude inhomogeneity increases. For the other nonlinearities however, this trend is entirely reversed. While all nonlinearities start at a similar level of amplitude inhomogeneity at the bifurcation, the sigmoid, periodic and clipped models all exhibit a decrease of $\delta(|x^*_i|)$ with $\beta$. The distributions become squeezed for high $\beta$ so that almost all of the spins become pinned to a level corresponding to the saturation levels in fig.\ref{fig2}b and amplitude inhomogeneity mostly vanishes. This is particularly pronounced for the clipped models, where the distribution is the narrowest of all the transfer functions. When comparing $\delta(|x^*_i|)$ with the success rate, we observe a clear correlation between the ability to find the ground state and the amount of amplitude inhomogeneity across the different models. As the inhomogeneity decreases, the implemented Ising model becomes closer to the target Hamiltonian and the ability to find the optimal solution is restored. Interestingly, while the amount of inhomogeneity is comparable between the different models, the success rate is not identical. Although the clipped model has an overall lower inhomogeneity, the success rate is highest for the periodic model. The lower success rate for the clipped model indicates that the spectrum of multistable excited states is different, either in the total number of states or in the size of their attractors. This shows that, while the suppression of inhomogeneity ensures the existence of the ground state at very similar values for $\beta$ for the different models in fig.\ref{fig7}, there can still be differences for the excited states. These differences are likely caused by the additional nonlinear terms in the Lyapunov function that are also discussed in previous sections and in the methods section. 

Overall, we find that the suppression of amplitude inhomogeneity leads to an overall improvement of the time-to-solution over the polynomial model across the different problems in fig.\ref{fig4}. Still, the computational performance advantage over the polynomial model only manifests itself for some of the problems. We attribute this to the varying difficulty of the different graphs contained in the Biq Mac library. For randomly generated graphs with spin numbers limited to $N=100$, there is a rather high probability of generating easy instances that can be solved in polynomial time \cite{KAL20}. We expect these easy instances to be more robust against faulty mapping due to amplitude inhomogeneity, while the differences in computational performance are more pronounced for difficult problems. To test this, we perform MaxCut benchmarks with graphs contained in the SuiteSparse Matrix Collection \cite{DAV11}. Compared to the Biq Mac library, the SuiteSparse Matrix Collection is a collection of sparse graphs with both unweighted ($J_{ij}=-1$) as well as bimodal edges ($J_{ij}=\left\{ -1,1\right\}$). The library contains both random and geometric topologies with spin numbers between $N=800$ and $N=5000$. Many of the instances contained in the SuiteSparse Matrix Collection are considered difficult and exact solutions are often not known. 

\begin{figure*}[tbp!]
	\centering
		\includegraphics[width=.99\textwidth]{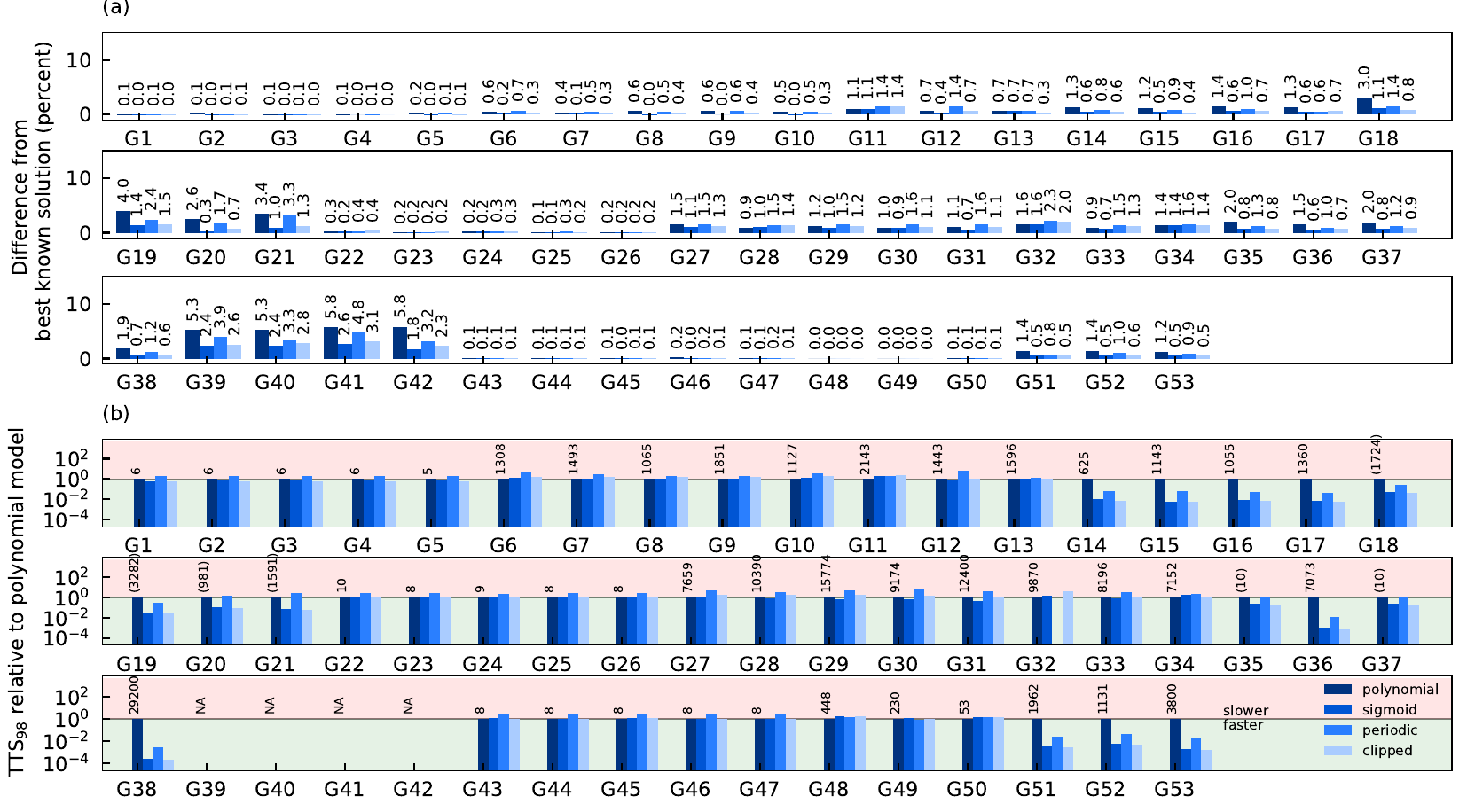}
	\caption{\textbf{Computational performance in SuiteSpare Matrix benchmark tasks} (a) Distance of the best solution obtained by the polynomial, sigmoid, periodic and clipped model from the best known solution for the SuiteSparse Matrix Collection benchmark tasks. (b) Time-to-solution of the periodic, sigmoid and clipped model relative to the polynomial model. The absolute value for the time-to-solution for the polynomial model is indicated by the numbers above the bars for the polynomial model. In cases where the polynomial model reaches only 95 percent of the best solution, $\mathrm{TTS_{95}}$ is shown instead (indicated by brackets around the TTS). Cases where the polynomial model fails to reach 95 percent of the best solution are indicated by $\mathrm{NA}$.}
	\label{fig5}
\end{figure*}

In fig.\ref{fig5}a, we show the relative distance $\Delta C= 100(1-C/C_{\mathrm{opt}})$ in percent of the best solution obtained by the different nonlinearities $C$ from the best known value reported in literature $C_{\mathrm{opt}}$ \cite{WAN19,MA17}. We find that all systems achieve solutions that are within just a few percent of or at the best known solution. Remarkably, this makes them comparable to state-of-the-art optimization methods such as simulated annealing or branch-and-bound algorithms without having to employ complex annealing schedules to further increase the solution quality. Considering the performance differences between the nonlinearities however, we can again observe that the polynomial model performs worse with an average distance of $\Delta C_{\mathrm{poly}}=1.3$ percent from the best known solution. The periodic and the clipped model achieve an average distances of $\Delta C_{\mathrm{per}}=1.1$ and $\Delta C_{\mathrm{clip}}=0.7$ respectively, while the best performance is achieved by the sigmoid nonlinearity with an average distance of $\Delta C_{\mathrm{sig}}=0.6$. This performance difference is especially striking for the set of bimodal problems with random connectivity and non-uniform node degree (G18, G19, G20, G21, G39, G40, G41, G42), where we observe improvements of up to four percent in the cut value over the polynomial model. For such bimodal problems, the probability of finding easy instances is significantly smaller than for unweighted graphs \cite{WAN19}, so that they can generally be assumed to be more difficult problems. We can thus observe a clear advantage in computational performance for such difficult problems that is gained by suppressing amplitude inhomogeneity through the nonlinear transfer function. 

This advantage is also reflected in the best time-to-solution obtained by the different models, which is shown in fig.\ref{fig5}b relative to the polynomial model (absolute values for all models are given in the supplementary tables 2 and 3). Since the ground state is not always reached for all problems, we consider the time-to-solution to reach 98 percent of the ground state $\mathrm{TTS}_{98}$. For instances where the solution of the polynomial model is more than 2 percent away from the best known solution, $\mathrm{TTS}_{95}$ is shown instead (indicated by brackets around the time-to-solution). Cases where the solution quality of the polynomial model is more than five percent away from the optimum are not considered in the following and indicated by $\mathrm{TTS}=\mathrm{NA}$ in fig.\ref{fig5}b. Similar to the Biq Mac library in fig.\ref{fig4}, we find that the polynomial model performs worse on average, while the sigmoid and the clipped model perform the best. For various problems, improvements of up to four orders of magnitude in the time-to-solution are obtained over the polynomial model. The largest differences are observed for graphs with a random connectivity and non-uniform node density, which can generally be considered to contain more difficult instances \cite{KAL20}. This again indicates a link between problem hardness and susceptibility to amplitude inhomogeneity. For uniform node densities and non-random connections on the other hand, which typically contain more easy instances \cite{KAL20}, we observe a lower susceptibility to amplitude inhomogeneity. 

\section*{Discussion}

We show how different gain-dissipative Ising machine designs can be unified in a single nonlinear feedback system that is fully described by three dimensionless parameters. Based on the generic pitchfork normal form, we describe how analog Ising spins can be generated by mimicking the general shape of the nonlinear transfer function of the polynomial model and discuss the performance of Ising machines based on periodic, sigmoid and clipped functions. By analyzing the Lyapunov function of the different nonlinear systems, we identify their ability to encode global energy minima of the Ising model as fixed points, whose stability is controlled by the linear gain. We find that different existing Ising machine concepts are in principal equally capable of implementing optimization problems and also demonstrate that sigmoid functions can be used as an alternative way of implementing analog spins that has not been considered in the context of Ising machines yet. Since the physical implementation of sigmoid functions has been investigated intensively as activation functions of artificial neurons, this creates an interesting link between Ising machines and neuromorphic computing concepts.

By performing benchmarks based on NP-hard MaxCut problems, we investigate the influence of the nonlinear transfer function on the quality of the solutions and the time to reach them. For both small and large-scale problems, we report significant differences in the computational performance for the different nonlinearities. While systems based on the pitchfork normal form may not be able to find the ground state, Ising machines using periodic, clipped and sigmoid nonlinearities offer better solution quality and a shorter time-to-solution for the same problems. Compared to the polynomial model, we observe improvements of up to four orders of magnitude in the time-to-solution and up to four percent in the solution quality relative to the optimal solution. With all systems evolving at the same dynamical timescale, we identify faulty mapping to the target Ising Hamiltonian as the cause for these performance differences. Due to this faulty mapping, local minima of the Ising Hamiltonian are stabilized while the ground state solution becomes destabilized. We link these mapping errors to amplitude inhomogeneity, which is caused by the analog nature of the spin system. Periodic, sigmoid and clipped transfer functions differ from the polynomial model in that they saturate for large gain. This causes squeezing of the amplitude distribution and reduces inhomogeneity as the gain is increased. We observe a direct correlation between this reduced inhomogeneity and the ability to find optimal solutions, which leads us to conclude that suppression of amplitude inhomogeneity through the transfer function can significantly aid in enhancing the computational performance of analog spin systems. 

This provides an intuitive explanation to some of the performance differences that have been reported for existing Ising machine concepts. In line with recent reports \cite{BOE19,TIU19,CHO19}, we find a clear computational advantage for systems with a saturable nonlinearity. The high sensitivity of computational performance to the nonlinear transfer function therefore strengthens the choice of such saturable nonlinearities for the design of analog Ising machines instead of systems based on pitchfork normal forms, while also motivating the search for other suitable nonlinear systems for future generations of Ising machines. Furthermore, saturable nonlinearities present an intriguing alternative to current approaches that aim to eliminate amplitude inhomogeneity by controlling the linear gain of each individual spin to force them to the same amplitude. While such systems have shown significant improvements in the solution quality \cite{LEL19,KAL18,LEL20,GOT21}, the necessity to control the gain of each spin creates a significant overhead and requires the addition of an active feedback system to the analog Ising machine. Using the transfer function to pin the spins to the same level instead provides a completely passive alternative that could retain the speed advantage of a fully analog system. This approach is also compatible with recently proposed annealing schemes that could further enhance the solution quality \cite{WAN19,PIE20a,MIL20}. Finally, beyond the considerations in this work, the sensitivity of computational performance of Ising machines to the shape of the transfer function could be further exploited to design nonlinear systems that are optimized for performance in specific optimization tasks. Combined with optical systems that are able to implement arbitrary nonlinear transfer functions \cite{WIL20,JHA20}, this would bring Ising machines closer to becoming fast and efficient accelerators for difficult optimization tasks.

\section*{Methods}

\subsection*{Condition for existence of fixed points}

In the following, we show the derivation of condition eq.\eqref{eqgain} both for homogeneous and inhomogeneous distributions of the fixed point amplitude $x^*_i$ for the polynomial model \cite{LEL17}. To consider both cases, we assume that $x^*_i$ can be written as $x^*_i=x\sigma_i$, where $x$ is the absolute value of the homogeneous spin amplitude ($x \geq 0$) and $\sigma_i$ is the spin state. To introduce inhomogeneity, we introduce an additional pre-factor $\nu_i$ for each spin so that $x^*_i=\nu_i x\sigma_i$. $\nu_i$ has a mean of $\bar{\nu_i}=1$ and follows a distribution with the standard deviation $\delta$. In case of a homogeneous amplitude distribution, $\delta=0$ and all $\nu_i$ are equal to one. The fixed points of the system can be found by setting the equation of motion equal to zero:
\begin{equation}
	0=(\alpha-1) \sigma_i\nu_i x-\sigma_i\nu_i^3x^3+\beta \sum_{j}J_{ij}\sigma_j\nu_j x \ \ .
\end{equation}
In the case of $x>0$, summation over all spins leads to:
\begin{equation}
	\frac{x^2}{N}\sum_i{\nu_i^2}=(\alpha-1) -\frac{\beta}{N} \sum_{ij}J_{ij}\sigma_i\sigma_j\frac{\nu_j}{\nu_i} \ \ .
\end{equation}
In order for a fixed point besides the trivial solution ($x=0$) to exist, the r.h.s. has to be larger than zero ($x>0$). This leads to the following inequality that describes the condition of existence for fixed points:
\begin{equation}
	\alpha-1 \geq \frac{\beta}{N} \sum_{ij}J_{ij}\frac{\nu_j}{\nu_i}\sigma_i\sigma_j \ \ . \label{eqgain2}
\end{equation}
In the case of homogenous spin amplitudes ($\nu_j=\nu_i$), this corresponds to the inequality in eq.\eqref{eqgain}. The right hand side of eq.\eqref{eqgain2} contains the Ising Hamiltonian \eqref{eq1} with the effective coupling matrix $J_{ij}'=J_{ij}\frac{\nu_j}{\nu_i}$. This shows that for a homogeneous amplitude distribution, the implemented Hamiltonian is equivalent to the target Ising Hamiltonian, since $J_{ij}'=J_{ij}$. For inhomogeneous amplitude distributions on the other hand, the implemented Ising Hamiltonian differs from the target Hamiltonian since every matrix element $J_{ij}$ is modified by a factor of $\frac{\nu_j}{\nu_i}$.

\subsection*{Approximation of the nonlinear transfer function for the sigmoid and periodic models}

To enable a direct comparison of the equations of motions for the periodic and sigmoid models against the polynomial model, eq.\eqref{eq5} and eq.\eqref{eq6} are approximated with polynomials. The polynomial approximation of the transfer function for the sigmoid model
\begin{equation}
	F_i(\{x_i\})=-x_i+\tanh(\alpha x_i + \beta\sum_j{J_{ij}x_j}) 
\end{equation}
follows from a third order Taylor expansion. Here, we consider a multi variable Taylor series for the spin amplitude of the isolated system $x_i$ and the spins injected by coupling with other spins $x_j$ for small values ($x_i \approx x_j \approx 0$). For simplicity, we consider the sum of $\sum_j{J_{ij}x_j}$ as a single variable. The resulting Taylor series to the third order is:
\begin{equation}
	F_i(\{x_i\}) \approx (\alpha-1)x_i + \beta\sum_j{J_{ij}x_j} -\frac{\alpha^3}{3}x_i^3 - \alpha^2\beta x_i^2\sum_j{J_{ij}x_j} - \alpha\beta^2x_i\left(\sum_j{J_{ij}x_j}\right)^2 - \frac{\beta^3}{3}\left(\sum_j{J_{ij}x_j}\right)^3 +O(x_ix_j^4)
\end{equation}
For the parameter scans of $\alpha$ and $\beta$, we assume that $0\leq \alpha,\beta\leq 1$. Furthermore, we consider the weak coupling regime where $\alpha \gg \beta$. This means that the terms containing $\beta$ contribute significantly less to the transfer function then the term only containing $\alpha$, since $\alpha^3 \gg \alpha^2\beta \gg \alpha\beta^2 \gg\beta^3$. While these higher order terms can cause a deviation of the Lyapunov function for the intended Ising Hamiltonian, we can assume that these deviations are small in the weak coupling regime and that the linear coupling term $\beta\sum_j{J_{ij}x_j}$ is dominant. For the final transfer function, we therefore neglect the third order terms in $x_i,x_j$ containing $\beta$. 
In a similar fashion, the Taylor expansion of the transfer function for the periodic model results in:
\begin{equation}
	F_i(\{x_i\}) \approx (\alpha-1)x_i + \beta\sum_j{J_{ij}x_j} -\frac{2\alpha^3}{3}x_i^3 - 2\alpha^2\beta 3 x_i^2\sum_j{J_{ij}x_j} - 2\alpha\beta^2 3x_i\left(\sum_j{J_{ij}x_j}\right)^2 - \frac{2\beta^3}{3}\left(\sum_j{J_{ij}x_j}\right)^3 \ \ .
\end{equation}
As for the sigmoid model, we neglect the third order terms in $x_i,x_j$ containing $\beta$. 

\subsection*{Numerical methods}

Simulations of the time evolution for the differential equations \eqref{eq4}, \eqref{eq5}, \eqref{eq6} and \eqref{eq7} are performed using the Euler method. For the simulations, a stepwitdth of $\Delta t=0.1$ is chosen. The number of total time steps is constant for all simulations and was chosen to be long enough to ensure converge to a steady state (3000 iterations for the BiqMac library, 5000 for the SuiteSparse Matrix library). At the bginning of each simulation, the system is initialized in the trivial fixed point $\{x_i\}=0$ and left to to evolve with $\alpha$ and $\beta$ at constant values during the entire evolution. The time to reach a given solution (e.g. the ground state) for each simulation is evaluated by tracking the Ising energy during the evolution and taking the point when the system first reaches the desired solution (either by converging or by a transient state). For the scans of $\alpha$ and $\beta$, the parameters were varied in the range $0 \leq \alpha,\beta\leq 1$. For each parameter point, the success rate and the time-to-solution were assessed from 50 independent simulations. 

\section*{Data availability} The authors declare that all relevant data are included in the manuscript. Additional data are available from the corresponding author upon reasonable request.

\section*{Acknowledgments}
We acknowledge financial support from the Research Foundation Flanders (FWO) under the grants G028618N, G029519N and G006020N as well as as the Hercules Foundation and the Research Council of the Vrije Universiteit Brussel. T.V.V. would like to thank R. Beausoleil for his mentorship.

\section*{Author Contributions} F.B. performed the simulations and analyzed the data. F.B., T.V.V., G.V. and G.V.d.S. discussed the results and wrote the paper.

\section*{Competing interest}
The authors declare no competing interest.


\section*{References}
\begin{thebibliography}{9}
\bibitem{THE17}
Theis, T. N.,  Wong, H.-S. P. The End of Moore's Law: A New Beginning for Information Technology. \textit{Computing in Science \& Engineering} \textbf{19}, 41-50 (2017).
\bibitem{STR19}
Strubell, E., Ganesh, A., McCallum, A. Energy and Policy Considerations for Deep Learning in NLP. \textit{Proceedings of the 57th Annual Meeting of the Association for Computational Linguistics} 1, 3645-3650 (2019).
\bibitem{AND15} Andrae, A., Edler, T. On Global Electricity Usage of Communication Technology: Trends to 2030. \textit{Challenges} \textbf{6}, 117-157 (2015).
\bibitem{JOH11}
Johnson, M. W. \textit{et al.} Quantum annealing with manufactured spins. \textit{Nature} \textbf{473}, 194-198 (2011).
\bibitem{YAM17a}
Yamamoto, Y. \textit{et al.} Coherent Ising machines- optical neural networks operating at the quantum limit. \textit{npj Quantum Information} \textbf{3}, 49 (2017).
\bibitem{DEN16}
Denchev, V. S. \textit{et al.} What is the computational value of finite-range tunneling? \textit{Physical Review X} \textbf{6}, 031015 (2016).
\bibitem{INA16a}
Inagaki, T. \textit{et al.} A coherent Ising machine for 2000-node optimization problems. \textit{Science} \textbf{354}, 603-606 (2016).
\bibitem{HAR16b}
Haribara, Y., Utsunomiya, S., Yamamoto, Y. Computational Principle and Performance Evaluation of Coherent Ising machine based on Degenerate Optical Parametric Oscillator Network. \textit{Entropy} \textbf{18}, 151 (2016).
\bibitem{SAL15}
Salath\'e, Y. \textit{et al.} Digital Quantum Simulation of Spin Models with Circuit Quantum Electrodynamics. \textit{Physical Review X} \textbf{5}, 021027 (2015).
\bibitem{BAR16}
Barends, R. \textit{et al.} Digitized Adiabatic Quantum Computing with a Superconducting Circuit. \textit{Nature} \textbf{534}, 222-226 (2016).
\bibitem{KIM10}
Kim, K. \textit{et al.} Quantum simulation of frustrated Ising spins with trapped ions. \textit{Nature} \textbf{465}, 590-593 (2010).
\bibitem{YAM16}
Yamaoka, M. \textit{et al.} A 20k-Spin Ising Chip to Solve Combinatorial Optimization Problems with CMOS Annealing. \textit{IEEE Journal of Solid-state Circuits} \textbf{51}, 303-309 (2016).
\bibitem{CHO19}
Chou, J., Bramhavar, S., Ghosh, S. Herzog, W. Analog Coupled Oscillator Based Weigthed Ising Machine. \textit{Scientific Reports} \textbf{9}, 14786 (2019).
\bibitem{UTS11}
Utsonomiya, S., Takata, K., Yamamoto, Y. Mapping of Ising models onto injection-locker laser systems. \textit{Optics express} \textbf{19}, 18091 (2011).
\bibitem{SHO17}
Shoji, T., Aihara, K., Yamamoto, Y. Quantum model for coherent Ising machines: Stochastic differential equations with replicator dynamics. \textit{Physical Review A} \textbf{96}, 053833 (2017).
\bibitem{BAB19}
Babaeian, M. \textit{et al.} A single shot coherent Ising machine based on a network of injection-locked multicore lasers. \textit{Nature Communications} \textbf{10}, 3516 (2019).
\bibitem{WAN13}
Wang, Z., Marandi, A., Wen, K., Byer, R. L., Yamamoto, Y. Coherent Ising machine based on degenrate optical parametric oscillators. \textit{Physical Review A} \textbf{88}, 063853 (2013).
\bibitem{MCM16}
McMahon, P. L. \textit{et al.} A fully programmable 100-spin coherent Ising machine with all-to-all connections. \textit{Science} \textbf{354}, 614-617 (2016).
\bibitem{BOE19}
B\"ohm, F., Verschaffelt, G., Van der Sande, G. A poor-man's coherent Ising machine based on opto-electronic feedback system for solving optimization problems. \textit{Nature Communications} \textbf{10}, 3538 (2019).
\bibitem{BER17}
Berloff, N. G. \textit{et al.} Realizing the classical XY Hamiltonian in polariton simulators. \textit{Nature Materials} \textbf{16}, 1120-1126 (2017).
\bibitem{TEZ20}
Tezak, N. \textit{et al.} Integrated Coherent Ising Machines Based on Self-Phase Modulation in Microring Resonators. \textit{IEEE Journal of Selected Topics in Quanutum Electronics} \textbf{26}, 5900115 (2020).
\bibitem{PIE19}
Pierangeli, D., Marcucci, G., Conti, C. Large-Scale Photonic Ising Machine by Spatial Light Modulation. \textit{Physical Review Letters} \textbf{122}, 213902 (2019).
\bibitem{PRA20}
Prahbu, M. \textit{et al.} Accelerating recurrent Ising machines in photonic integrated circuits. \textit{Optica} \textbf{7}, 551 (2020).
\bibitem{OKA20}
Okawachi, Y. \textit{et al.} Demonstration of chip-based coupled degenerate optical parametric oscillators for realizing nanophotonic spin-glass. \textit{Nature Communications} \textbf{11}, 4119 (2020).
\bibitem{LEL19}
Leleu, T., Yamamoto, Y., McMahon, P., Aihara, K. Destabilization of Local Minima in Analog Spin Systems by Correction of Amplitude Heterogeneity. \textit{Physical Review Letters} \textbf{122}, 040607 (2020).
\bibitem{KAL18}
Kalinin, K., Berloff, N. G. Global optimization of spin Hamiltonians with gain-dissipative systems. \textit{Scientific Reports} \textbf{8}, 17791 (2018).
\bibitem{TIU19}
Tiunov, E. S., Ulanov, A. E., Lvovsky, A. I. Annealing by simualting the coherent Ising machine. \textit{Optics Express} \textbf{27}, 10288 (2019).
\bibitem{GOT19}
Goto, H., Tatsumura, K., Dixon, A. R. Combinatorial optimization by simulating adibatic bifurcations in nonlinear Hamiltonian systems. \textit{Science Advances} \textbf{5}, eaav2372 (2019).
\bibitem{LEL17}
Leleu, T., Yamamoto, Y., Utsunomiya, S., Aihara, K. Combinatorial optimization using dynamical phase transitions in driven-dissipative systems. \textit{Physical Review E} \textbf{95}, 022118 (2017).
\bibitem{BOE18}
B\"ohm, F. \textit{et al.} Understanding dynamics of coherent Ising machines through simulation of large-scale 2D Ising models. \textit{Nature Communications} \textbf{9}, 5020 (2018).
\bibitem{INA16}
Inagaki, T. \textit{et al.} Large-scale Ising spin network based on degenerate parametric oscillators. \textit{Nature Photonics} \textbf{10}, 415-419 (2016).
\bibitem{LUC14}
Lucas, A. Ising formulation of many NP-hard problems. \textit{Frontiers in physics} \textbf{2}, 1-15 (2014).
\bibitem{KUE11}
Kuehn, C. A mathematical framework for critical transitions: Bifurcations, fast-slow systems and stochastic dynamics. \textit{Physica D: Nonlinear Phenomena} \textbf{240}, 1020-1035 (2011).
\bibitem{KAN21}
Kanao, T., Goto, H. High-accuracy Ising machine using Kerr-nonlinear parametric oscillator with local four-body interaction. \textit{npj Quantum Information} \textbf{7}, 18 (2021).
\bibitem{HOP85}
Hopfield, J. J., Tank, D. W. "Neural" computation of decisions in optimization problems. \textit{Biological cybernetics} \textbf{52}, 141-152 (1985).
\bibitem{MIS18}
Miscuglio, M. \textit{et al.} All-optical nonlinear activation function for photonic neural networks. \textit{Optical Materials Express} \textbf{8}, 3851 (2018).
\bibitem{WIL20}
Williamson, I. A. D. \textit{et al.} Reprogrammable Electro-Optic Nonlinear Activation Functions for Optical Neural Networks. \textit{IEEE Journal of Selected Topics in Quantum Electronics} \textbf{26}, 7700412 (2020).
\bibitem{JHA20}
Jha, A., Huang, C., Prucnal, P. R. Reconfigurable all-optical nonlinear activation function for neuromorphic photonics. \textit{Optics Letters} \textbf{45}, 4819 (2020).
\bibitem{MOU19}
Mourgias-Alexandris, G. \textit{et al.} An all-optical neuron with sigmoid activation function. \textit{Optics Express} \textbf{27}, 9620 (2019).
\bibitem{LU00}
Lu, C., Shi., B., Chen, L. Analogue circuit realization of a programmable sigmoidal function and its derivative for on-chip BP learning. \textit{IEEE APCCAS 2000. 2000 IEEE Asia-Pacific Conference on Circuits and System. Electronic Communication Systems (Cat. No.00EX394)}, 626-629 (2000). 
\bibitem{TSM19}
Tsmots, I., Skorokhoda, O., Rabyk, V. Hardware Implemenation of Sigmoid Activation Functions using FPGA. \textit{2019 IEEE 15th International Conference on the Experience of Designing and Application of CAD Systems (CADSM)}, 34-38 (2019).
\bibitem{REN07}
Rendl, F., Rinaldi, G., Wiegele, A. A Branch and Bound Algorithm for Max-Cut Based on Combining Semidefinite and Polyhedral Relaxations. In \textit{Integer Programming and Combinatorial Optimization}, vol. 4513 LNCS, 295-309 (Springer Heidelberg, Berlin, Heidelberg, 2007).
\bibitem{ROQ20}
Roques-Carmes, C. \textit{et al.} Heuristic recurrent algorithm for photonic Ising machines. \textit{Nature Communications} \textbf{11}, 249 (2020).
\bibitem{PIE20}
Pierangeli, D., Marcucci, G., Brunner, D., Conti, C. Noise-enhanced spatial-photonic Ising machine, \textit{Nanophotonics} \textbf{9}, 4109-4116 (2020).
\bibitem{GAR90}
Garey, M., Johnson, D. S. \textit{Computers and Intractability: A Guide to the Theory of NP-Completeness} (W. H. Freeman \& Co, New York, NY, USA 1990).
\bibitem{KAL20}
Kalinin, K., Berloff, N. G. Complexity continuum within Ising formulations of NP problems. Preprint at http://arxiv.org/abs/2008.00466 (2020).
\bibitem{DAV11}
Davis, T. A., Hu, Y. The university of Florida sparse matrix collection. \textit{ACM Transactions on Mathematical Software} \textbf{38}, 1-25 (2011).
\bibitem{WAN19}
Wang, T., Roychowdhurry, J. OIM: Oscillator-Based Ising Machines for Solving Combinatorial Optimization Problems. In \textit{Lecture Notes in Computer Science (including Lecture Notes in Artificial Intelligence and Lecture Notes in Bioinformatics)}, vol. 11493 LNCS, 232-256 (2019).
\bibitem{MA17}
Ma, F., Hao, J.-K. A multiple search operator heuristic for the max-k-cut problem. \textit{Annals of Operations Research} \textbf{248}, 365-403 (2017).
\bibitem{LEL20}
Leleu, T. \textit{et al.} Scaling advantage of nonrelaxational dynamics for high-performance combinatorial optimization. Preprint at http://arxiv.org/abs/2009.04084 (2020).
\bibitem{GOT21}
Goto, H. \textit{et al.} High-performance combinatorial optimization based on classical mechanics. \textit{Science Advances} \textbf{7}, eabe 7953 (2021).
\bibitem{PIE20a}
Pierangeli, D., Marcucci, G., Conti, C. Adiabatic evolution on a spatial-photonic Ising machine. \textit{Optica} \textbf{7}, 1535 (2020).
\bibitem{MIL20}
Mills, K., Ronagh, P., Tamblyn, I. Finding the ground state of spin Hamiltonians with reinforcement learning. \textit{Nature Machine Intelligence} \textbf{2}, 509-517 (2020).

\end{thebibliography}
\end{document}